\documentclass[pre,preprint,showpacs]{revtex4} 
\input{epsf}  
\usepackage{graphicx}
\usepackage{float}
 
\begin{document}  
\title{Similar is Better: Speed Variability Reduces Traffic Flow }
  
\author{Azi Lipshtat }  
\affiliation{  
Department of Pharmacology and Systems Therapeutics
Mount Sinai School of Medicine
New York, NY 10029}  

\begin{abstract}  
Every driver knows that severe weather conditions cause traffic congestions. In many cases there is no direct reason for the congestion, and people tend to attribute it to the slow driving mode. Our computational study shows that the slow driving cannot account for this phenomenon. It comes out that the reason is not the decrease in mean velocity, but rather the change in the speed distribution on the road. Width of the distribution, namely - the number of "very fast" or "very slow" vehicles, affects both the need and the availability of lane change and bypassing. Besides providing insight and analyzing the underlying mechanism of a collective phenomenon, this example sheds light on a fundamental aspect of computational modeling. Although simple-as-possible models are desirable and provide better understanding of the real important component of the modeled system, they can very easily turn into over-simplifications and miss relevant qualitative phenomena.   
\end{abstract}

\maketitle

\section{Introduction}

Traffic problems have been modelled mathematically and computationally over the last two decades in various contexts. There is a large variety of models, for one dimensional \cite{Bando,Paul,Schreckenberg}, two dimensional \cite{Biham,Chung}, or network geometries \cite{Schreckenberg2,Youn,Liang}, and for either single-\cite{Schreckenberg,Hayakawa} or multi-lane \cite{Davis,Helbing} roads. However, all these models assumed similar speed for all vehicles, or at least determined the speed according to the same set of deterministic rules. In reality, of course, each driver has a different policy in determining the vehicle velocity in response to any given set of conditions. The amount of variability between drivers is not known and its effects have not yet been studied. Speed variability causes drivers to move between lanes. The effect of changing lanes on multi-lane traffic flow has been modelled and analyzed in the past \cite{Nagel,Rickert,Laval}. It was shown that voids, caused by changing lanes, reduce the total traffic flow \cite{Laval}. On the other hand, under realistic conditions, where not all vehicles run in the same speed, changing lanes is crucial to maintaining the flow. Without the option of changing lanes, the slowest vehicle would determine the speed for all other vehicles on the lane. Thus, changing lanes can either decrease or increase the average speed.  Since the need for lane changing emerges from the heterogeneity in speeds, we wanted to examine systematically the effect of speed distribution on traffic flow. To this end we constructed a simple model of multi-lane road with distribution of vehicles' speed. In the next Section we present our computational model in details. Results and analysis are presented in Section \ref{Sec:results}. We conclude with a general discussion about modeling of complex systems in Section \ref{Sec:discussion}.

\section{Computational Model}
\label{Sec:model}

Our model consists of continuous single- or multi-lane freeway. The dynamics is deterministic and the only stochastic dynamic component is the time delay between arrivals of new vehicles onto the freeway. Movement of a vehicle on the road is determined by simple rules: Each vehicles has a desired velocity $v_d$. This is the velocity of the vehicle under ideal conditions, where there are no other vehicles on the road. The desired velocities are taken from a normal distribution \cite{Partha} with mean $V$ and standard deviation $\sigma=C_v\times V$, where $C_v$ is the coefficient of variance. Presence of other vehicles in front of a running car may cause deceleration in order to avoid accidents. In case that the actual velocity, $v_a$, is slower than the desired velocity, and moving to a neighboring lane would improve the vehicle position, then the vehicle changes its lane. 

A numerical code was written, using Matlab$^{\rm TM}$ \cite{Matlab}. The initial condition was an empty highway, consists of $L$ semi-infinite lanes. A list of desired velocities was taken from a normal distribution with given mean and standard deviation as indicated in each case. The negative velocities have been deleted and replaced by new velocities, taken from the same distribution. New vehicles entered a lane at $x=0$, conditioned that last car in that lane was far enough so that there will be no accident. Then a time gap to arrival of the next vehicle to that lane was randomly taken from uniform distribution between $0$ and $T$ (sec.). Unless otherwise indicated, we used T=2 sec. Each of the iterations was composed of three steps: speed update, driving, and lane changes.  In the first step, vehicles with $v_a < v_d$ whose distance to the next vehicle was large enough, increased their velocity. If the distance to next vehicle was greater than 5 vehicle lengths, the velocity was updated to $v_d$, and in case of smaller distances, the increase was in a lesser extent. Then, vehicles that were too close to the next vehicle reduced their speed to the exact velocity required to avoid accidents at the next time step. Then all vehicles updated their location according to their respective velocities.  At the bypassing step, there were two necessary conditions for a vehicle in order to change its lane: (1) running too slow, namely $v_a < 95\%\, v_d$; and (2) having better position at a neighbor lane. Better position means either larger free space to the next vehicle or faster vehicle in front. All vehicles that fulfilled these two conditions changed their lanes. The results have been averaged over 40 runs of the simulations, where each run simulated one hour of traveling vehicles.

\section{Results and Analysis}
\label{Sec:results}

We define a satisfaction index $\eta$ by calculating the ratio $v_a/v_d$, averaged over all running vehicles. $\eta =1$ implies ideal flow where all vehicles are running in their desired speed. Lower values of $\eta$ mean that there is a slow-down of the traffic and vehicles are forced to run slower than they would like to. In other words,  $\eta$ is a measure of traffic congestion.
In the extreme case of $\sigma=0$, all vehicles have the same desired velocity $v_d$. In this situation, there is no reason for any vehicle to slow down, since the vehicle in front of any given vehicle does not run slower than the vehicle behind. Thus, in this situation $v_a=v_d$ for all vehicles and there is perfect satisfaction, namely $\eta=1$. As the variability in speeds gets larger, there are more slow vehicles which force the cars behind them to run slower than they want to, and the satisfaction index decreases. In a single lane free way, there is no way to bypass the slow vehicles, and thus $\eta$  decreases rapidly with $C_v$. This decrease is not dependent on the average velocity V but only on the width of the distribution (Figure 1A).

Opening more lanes enables bypassing and improves the traffic flow significantly. Does this improvement depend on the speed distribution or on the average velocity? Numerical simulations show that the width of the distribution has a much more significant role than the absolute velocities (Figure 1B). Increase of the coefficient of variance causes a decrease in the efficiency of the traffic flow, and this decrease is similar for a broad range of mean velocities. Increasing the variance leads to two opposing effects: on the one hand, higher variability in velocities means more cases in which fast vehicles are significantly delayed by slower ones. Thus, as the variability increases, there is more need for bypassing and lane changing. On the other hand, more variability leads also to larger distances between successive vehicles, making the lane changing easier. Changes in the width of distribution can thus change the road dynamics and it is not surprising to see that the traffic flow is dependent on $C_v$. However, changing the mean velocity without changing $C_v$ is mathematically equivalent to changing the time (or length) units. This is similar to taking a film of a road with high mean velocity and presenting it in slow motion. Obviously, if there is no congestion in the fast running, there cannot be either in the slow one.

The satisfaction index $\eta$  may be affected by the number of possible lanes. One may hypothesize that more lanes will result in more efficient flow. To some extent, this is indeed the case. A significant improvement is observed by changing a single lane road into a 2-lanes way. However, the contribution of any additional lane is minor (Figure 2). Below we will show how this observation can be useful under certain circumstances.

Since in our model vehicles are added to each lane independently, the number of lanes does not change the global road density. However, vehicle density is a parameter of the system which can be tuned. This parameter may affect the traffic flow. As density increases, the average space between vehicles decreases and it becomes more difficult to change lanes. However, high density is the situation where lane change is mostly required in order to maintain high traffic flow. Thus, in addition to the distribution width, the total density is also a crucial parameter. In our model, the density is determined by the time gap between arrivals of new vehicles to the road. Very small time gaps are not feasible, since a vehicle cannot enter the road if the car in front is too close. At the other extreme, a very large gap (and low density) implies that there is almost no interaction between vehicles and the actual velocity of any vehicles is the same as its desired speed. Between these limit cases there is an increase in the satisfaction index   as function of the time gap $T$ (Figure 3). 

Interestingly, not all drivers respond in a similar way to changes in the road conditions such as bad weather or traffic congestions. In a situation where all vehicles slow down and reduce their speed by the same factor, the coefficient of variation should remain unchanged. However, observations show that this is not the case, and $C_v$ changes with changes in the traffic flow. From measurements of real vehicles speeds, taken by the Washington State Department of Transportation \cite{WSDoT}, it can be easily seen that the variation increases dramatically at rush hours, together with the number of vehicles (Figure 4). This indicates that drivers may act differently under different conditions. The different responses  change the speed distribution and thus affect the congestion.

According to the US Federal Highway Administration, $15\%$ of the congestions are caused by weather \cite{FHWA}. Bad weather may cause congestion by many indirect ways - causing floods and accidents, opening pits, and providing other reasons for lane closures. However, every driver knows from self experience that congestions are prevalent under bad weather conditions even without any special events. Many drivers tend to drive slowly in rain, but this fact by itself cannot be account for congestion, as we showed above. Our results suggest that a possible reason for congestion formation is not the slow-down per se, but rather a change in the distribution. The rate of slowing down is different from one driver to another - some drivers do not change their speed whereas others decrease it significantly. As a result, the speed distribution gets broader, and this change decreases the traffic flow efficiency. To verify this hypothesis, we have compared the speed distribution of a sunny day to that of a rainy one. Speed measurements were taken at the same location and at the same day of the week, and the total number of vehicles was similar. As shown in Figure 5, the variation under rain condition is much higher and lasts for more hours than in a clear day. The wide distribution explains the high frequency of traffic congestions under bad weather conditions, even in cases there is no direct obstacle.

This observation suggests a simple method for improving traffic flow under conditions which are known to cause wide velocity distribution. All that it takes is dividing a multi-lane highway into groups of 2 or 3 lanes each, and directing vehicles to a specific group based on their speed. By doing so, we effectively create several parallel mini highways with narrow distribution in each of them. Since we showed above that the number of lanes is not a major parameter, the increase in efficiency due to narrowing the distribution is higher than the loss due to the reduction in number of lanes.

\section{Discussion}
\label{Sec:discussion}

When modeling collective behavior, such as traffic flow, evolution of populations, or financial dynamics, one should always look for the balance between two opposing arguments. On the one hand, we tend to view the global variables as reliable descriptions not only of the population but also of the individuals within the population - the mean behavior is the behavior of the average individual. This is a simplifying argument which helps developing and analyzing models of complex systems. On the other hand, these models may miss important aspects of the system which are either driven by the intra-population variability and diversity or simply averaged out and can be viewed only in local observations \cite{Lipshtat, Loinger}. It was Albert Einstein who stated that "It can scarcely be denied that the supreme goal of all theory is to make the irreducible basic elements as simple and as few as possible without having to surrender the adequate representation of a single datum of experience" \cite{Einstein}. This statement is usually quoted as "A theory should be as simple as possible, but no simpler". The same can be said about computational modeling of complex systems. A good model should include as few assumptions, constraints, or details as possible, and yet be able to capture and simulate the essence of the real system. This is the reason why, when possible, many modelers ignore the existence of variability in their parameters and refer to an average value only. Though such models can be valid and describe well many systems, there are other cases where the deviation from the average is the driving force for quantitative and qualitative phenomena. Ignoring the variability in these cases may result in misleading conclusions. Variability was shown as a determinant factor in ecologic systems \cite{Clark}, neurology \cite{Eggermont,Murphy}, social studies \cite{Durdu} and other fields \cite{Gould}. Ignoring it may be justified in certain cases, but taking it into account can provide insights and explain quantitative and qualitative phenomena.

\subsection*{Acknowledgments} 
I would like to thank Ravi Iyengar for his help. Thanks also to Jim Hawkins
and Cynthia Whaley from the Washington State Department of Transportation for providing the data.
This research is supported by NIH grant P50 - GM071558  Systems Biology Center grant.


\newpage

\begin{figure}[h]
\vspace{-3cm}
\includegraphics[width=15cm]{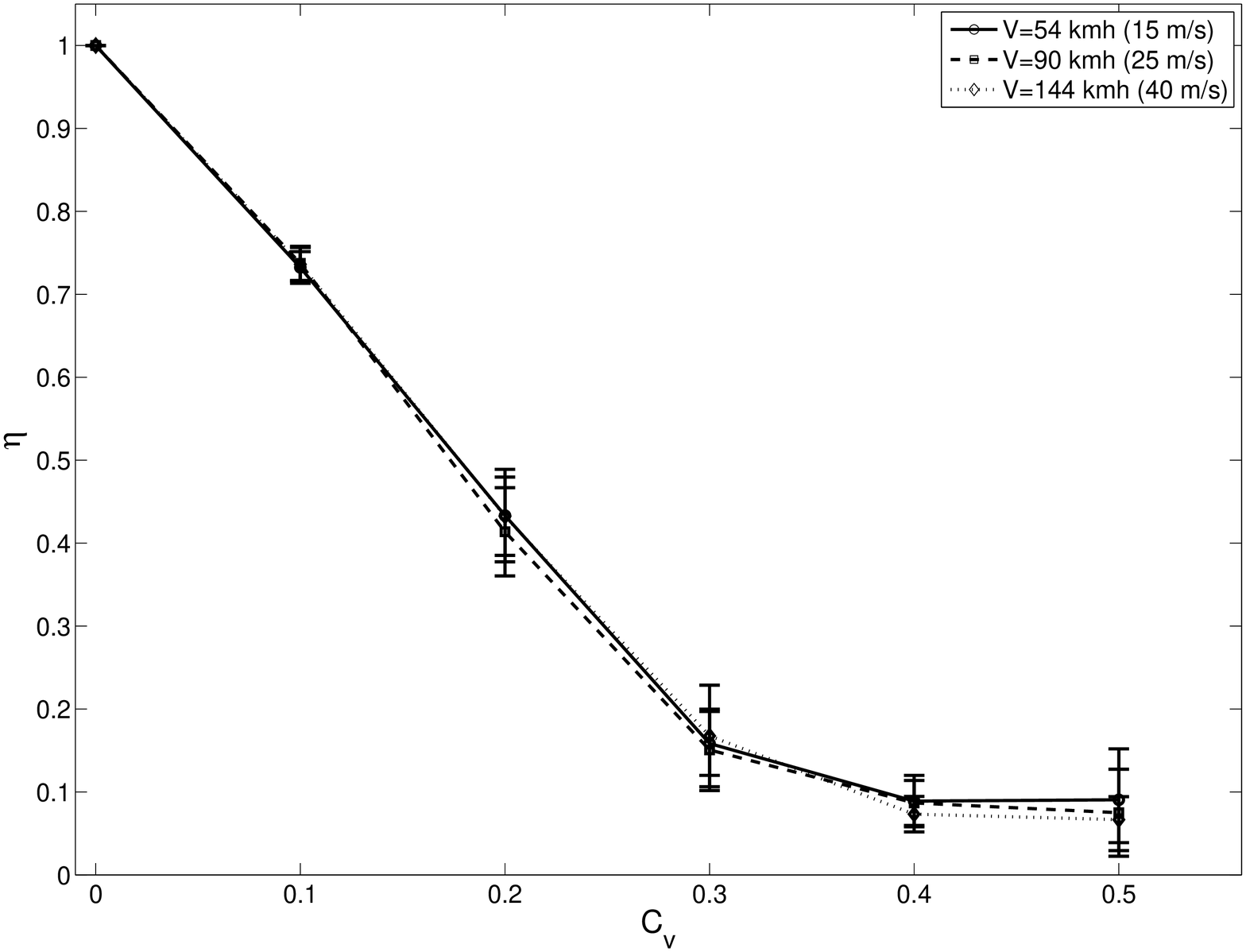}
\vspace{-1cm}
\includegraphics[width=15cm]{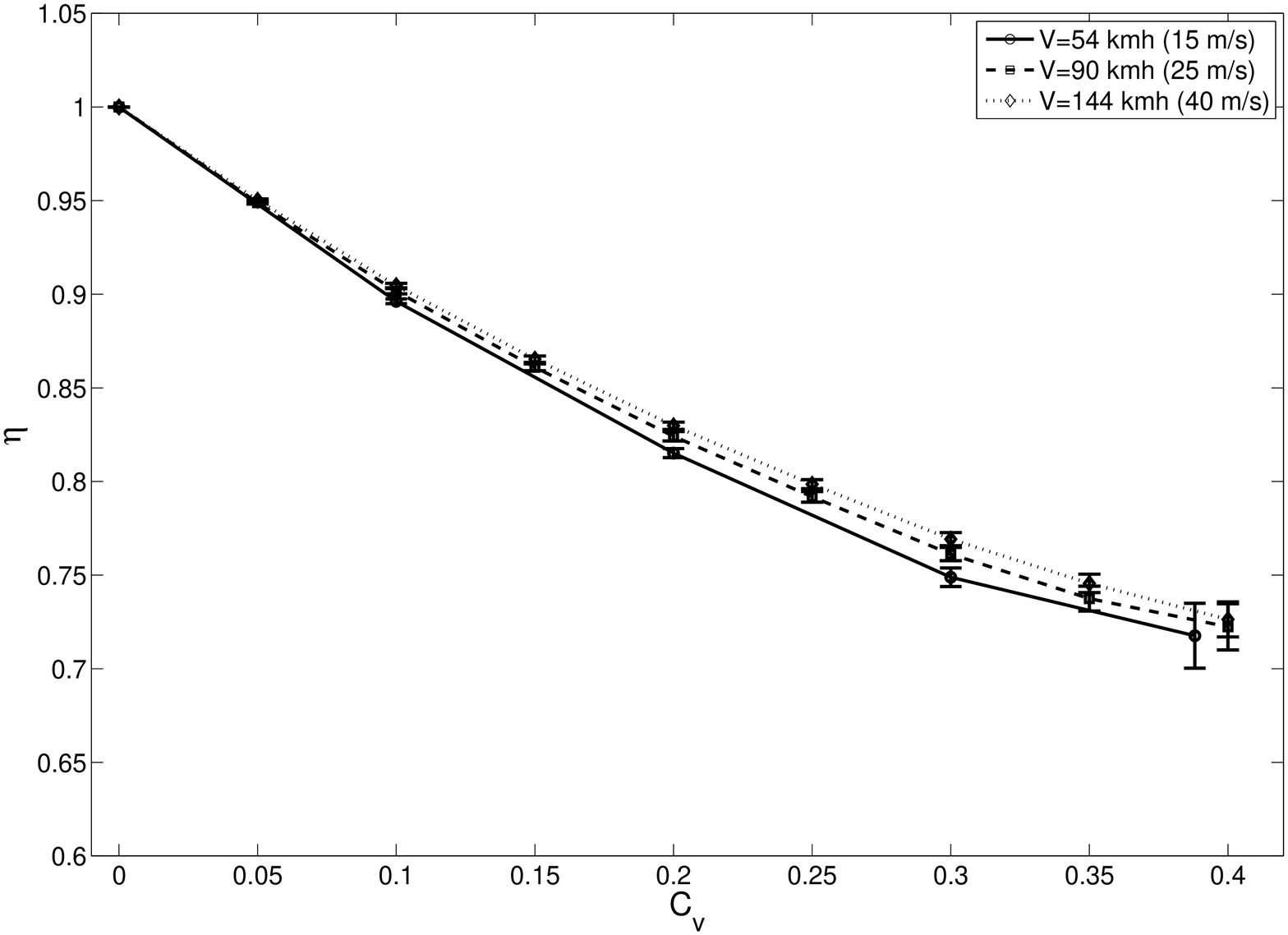}
\label{Fig:1}
\caption{The satisfaction index $\eta$ depends on the variance coefficient but not on the mean velocity. (A). Effect of variance coefficient in case of a single lane. The satisfaction index decreases rapidly with variance. At lower values of $C_v$, by increasing $C_v$ we have more slow vehicles (as well as fast ones). Since there are no negative velocities in this model, for $C_v > 0.4$ increasing the variance means addition of fast vehicles only. Since these vehicles cannot run on their desired velocity, changing of $C_v$ has no effect. (B). In a 5-lanes highway, bypassing and lane changing enable faster traffic flow. However, high variability still causes obstacles and reduces satisfaction. }
\end{figure}

\begin{figure}[h]
\label{Fig:2}
\includegraphics[width=18cm]{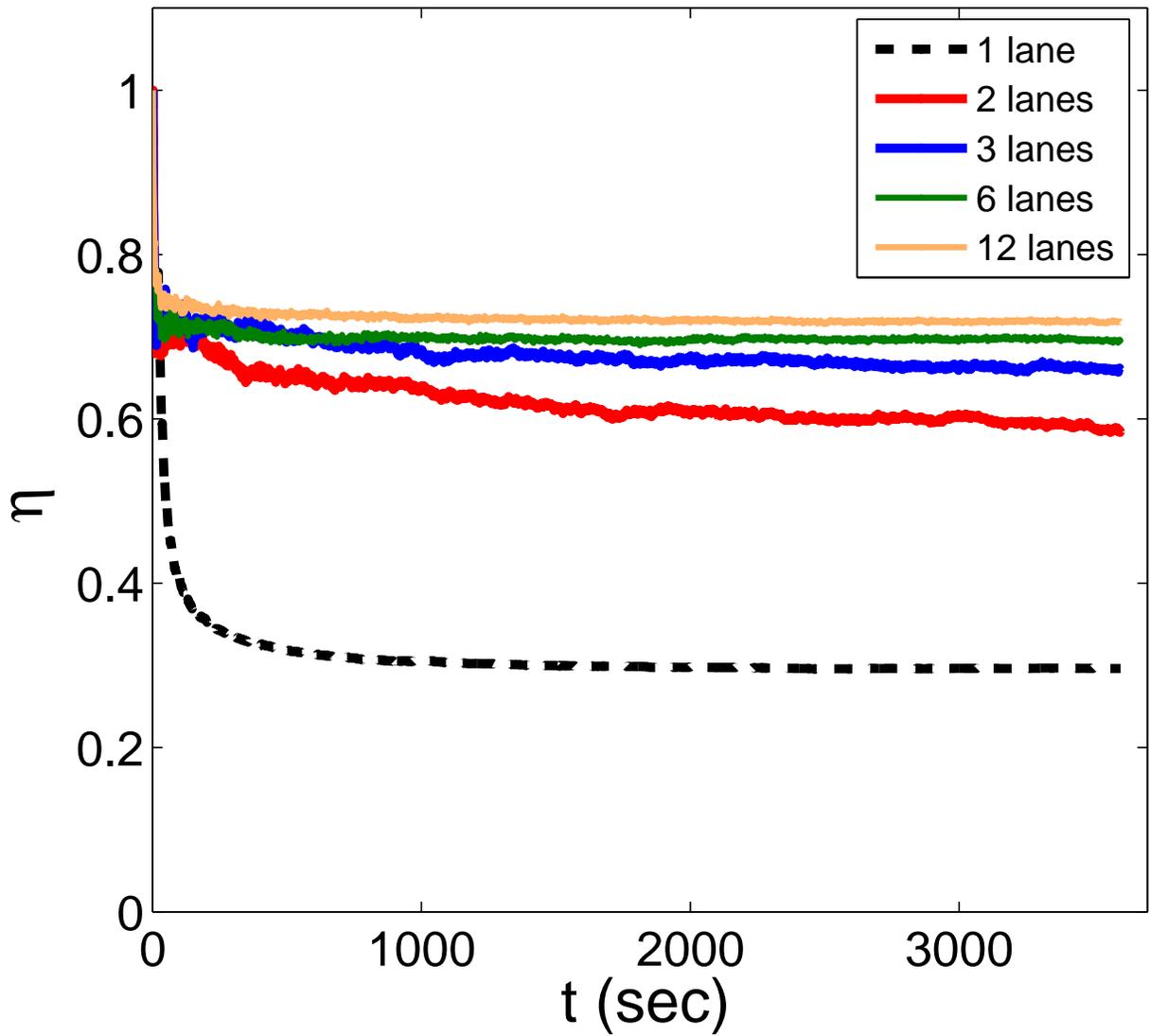}
\caption{As the number of lanes increases, satisfaction index increases as well. However, the effect of addition lanes beyond the second is minor. (Single simulation for each number of lanes. $C_v=0.25$)}
\end{figure}

\begin{figure}[h]
\label{Fig:3}
\includegraphics[width=18cm]{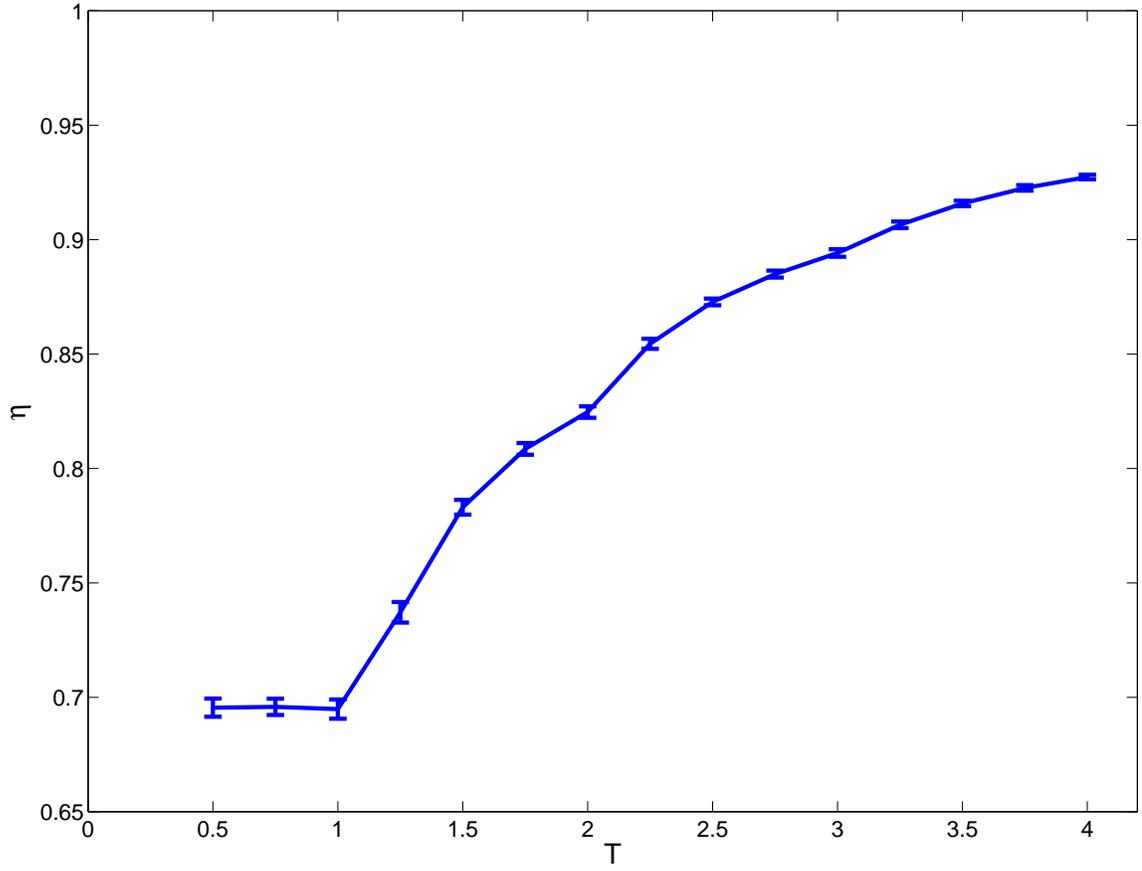}
\caption{Maximal time delay between arrivals of successive vehicles affects positively the satisfaction index. The actual time gap is taken randomly from uniform distribution between 0 and $T$.}
\end{figure}

\begin{figure}[h]
\label{Fig:4}
\includegraphics[width=18cm]{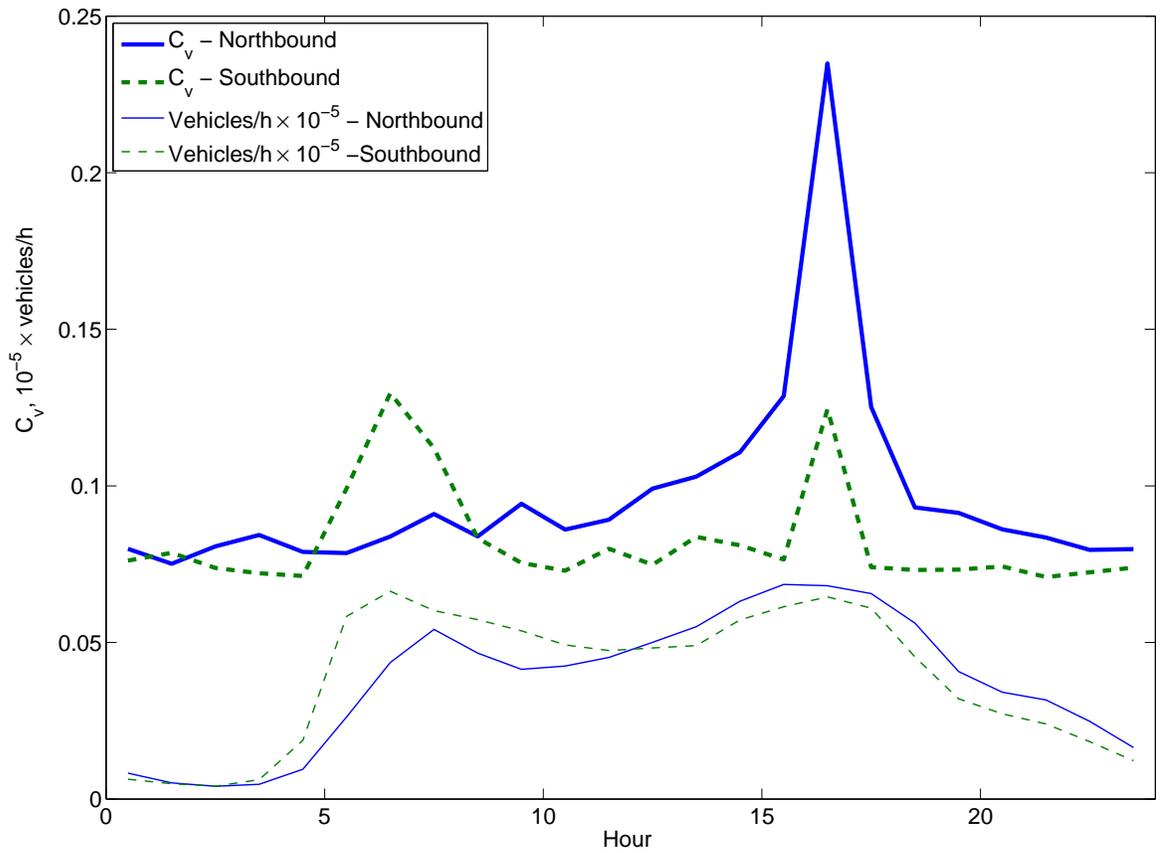}
\caption{Coefficient of variation of speed distribution at each time of the day. Measurements were taken by the Washington State Department of Transportation on May 27$^{\rm th}$ 2008, at the interstate road I-5, near mile post 185. There are 4 lanes to each direction and the legal speed limit is 60 mph (about 96 kmh).}
\end{figure}

\begin{figure}[h]
\label{Fig:5}
\vspace{-3cm}
\includegraphics[width=15cm]{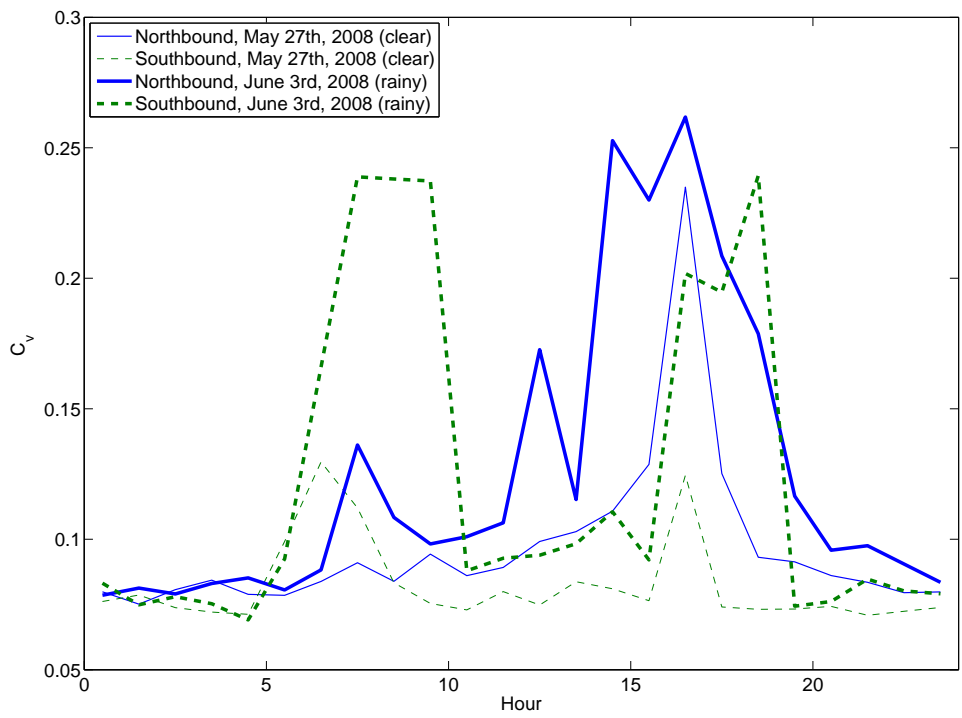}
\vspace{-1cm}
\includegraphics[width=15cm]{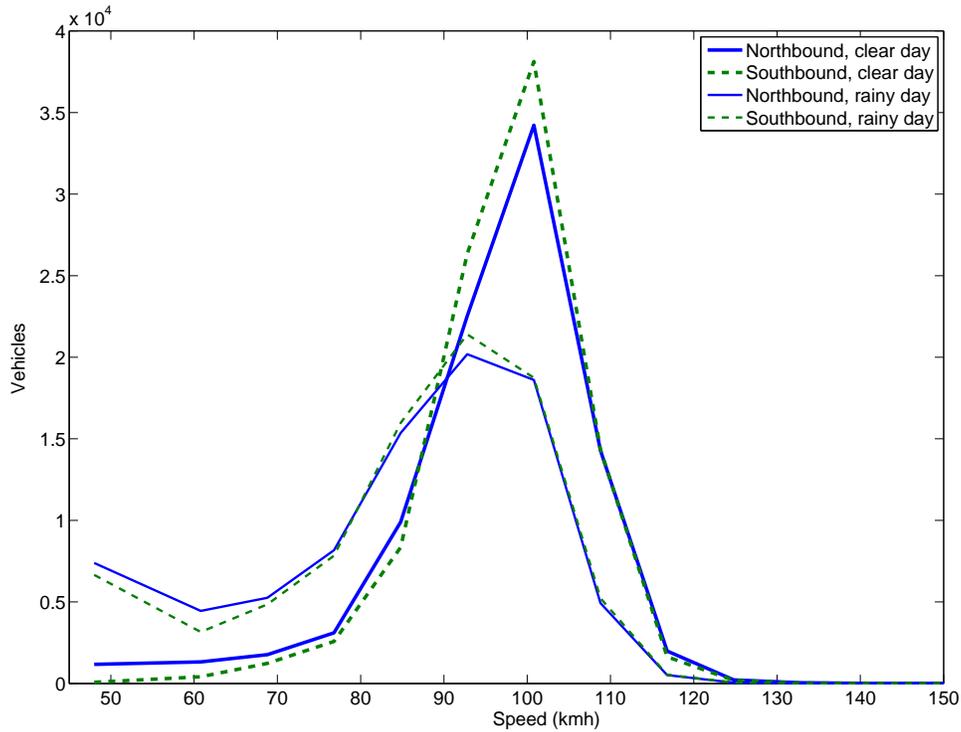}
\caption{Comparison of speed distributions at clear and rainy days. (A) Coefficient of variation. In clear days there is high variation at rush hours only, whereas in a rainy day the variation is higher and may occur at any time of the day. Both measurements were taken at the same location, on two successive Tuesdays. (B) Speed distribution of the whole day. The total number of vehicles was similar in two days (90571 (92833) vehicles went northbound (southbound) on May 27$^{\rm th}$, and 84913 (84456) on June 3${\rm rd}$), but the distribution is different. Weather information was taken from the National Oceanic and Atmospheric Administration \cite{noaa}.
 }
\end{figure}

\end{document}